\documentclass[sigconf]{acmart}
\AtBeginDocument{%
  \providecommand\BibTeX{{%
    \normalfont B\kern-0.5em{\scshape i\kern-0.25em b}\kern-0.8em\TeX}}}

\usepackage{times}
\usepackage{url}
\usepackage{graphicx}
\usepackage{booktabs}
\usepackage{latexsym}
\usepackage{xspace}
\usepackage[position=bottom]{subfig}

\renewcommand{\comment}[1]{}
\newcommand{\name}{\textsc{VerSaChI}\xspace}
\newcommand{\newsym}[1]{\ensuremath{\mathcal{#1}}}

\newcommand{\graph}[1]{\ensuremath{\mathcal{#1}}}

\newcommand{\ilg}{\ensuremath{IL_\graph{G}}\xspace}
\newcommand{\lnlg}{\ensuremath{LNL_\graph{G}}\xspace}
\newcommand{\ilq}{\ensuremath{IL_\graph{Q}}\xspace}
\newcommand{\lnlq}{\ensuremath{LNL_\graph{Q}}\xspace}
\newcommand{\lcvg}{\ensuremath{LCV_\graph{G}}\xspace}
\newcommand{\lcvq}{\ensuremath{LCV_\graph{Q}}\xspace}
\newcommand{\vpm}{\ensuremath{CP}\xspace}

\newcommand{\match}[1]{\ensuremath{Match^{(#1)}}\xspace}

\copyrightyear{2021}
\acmYear{2021}
\setcopyright{acmlicensed}
\acmConference[CIKM '21]{Proceedings of the 30th ACM International Conference on Information and Knowledge Management}{November 1--5, 2021}{Virtual Event, QLD, Australia}
\acmBooktitle{Proceedings of the 30th ACM International Conference on Information and Knowledge Management (CIKM '21), November 1--5, 2021, Virtual Event, QLD, Australia}
\acmPrice{15.00}
\acmDOI{10.1145/3459637.3482217}
\acmISBN{978-1-4503-8446-9/21/11}

\acmSubmissionID{rgsp2563}

\newcommand{\ab}[1]{\textcolor{red}{#1}}

\begin{document}
\fancyhead{}

\title{\name: Finding Statistically Significant Subgraph Matches using Chebyshev's Inequality}

\author{Shubhangi Agarwal}
\affiliation{%
  \institution{Indian Institute of Technology Kanpur}
  \country{India}
}
\email{sagarwal@cse.iitk.ac.in}

\author{Sourav Dutta}
\affiliation{%
  \institution{Huawei Research Centre}
  \city{Dublin}
  \country{Ireland}
}
\email{sourav.dutta2@huawei.com}

\author{Arnab Bhattacharya}
\affiliation{%
  \institution{Indian Institute of Technology Kanpur}
  \country{India}
}
\email{arnabb@cse.iitk.ac.in}

\begin{abstract}
	Approximate subgraph matching, which is an important primitive for many
	applications like question answering, community detection, and motif
	discovery, often involves large labeled graphs such as knowledge
	graphs, social networks, and protein sequences.  Effective methods for
	extracting matching subgraphs, in terms of label and structural
	similarities to a query, should depict accuracy, computational efficiency,
	and robustness to noise.  In this paper, we propose \name for finding the
	top-k most similar subgraphs based on 2-hop label and structural overlap
	similarity with the query. The similarity is characterized using Chebyshev's 
	inequality to compute the chi-square statistical significance for measuring 
	the degree of matching of the subgraphs. Experiments on real-life graph 
	datasets showcase significant improvements in terms of accuracy compared to 
	state-of-the-art methods, as well as robustness to noise.
\end{abstract}



\maketitle

\section{Introduction}
\label{sec:intro}

With the growth of Open Linked Data in the form of knowledge graphs, social
networks, bioinformatic 
structures, and road networks, 
efficient graph mining poses a challenging 
problem~\cite{www1,chi80}.
Such large data sources are represented as labeled graphs, where entities are
modeled as vertices, while their relationships 
are captured by
edges, with labels defining the attributes of entities and relations. 
{\em Subgraph querying} is used 
across several domains including
frequent pattern search in data mining~\cite{c49}, community detection in IR~\cite{gf2}, 
question answering in NLP~\cite{gf4}, object
recognition in computer vision~\cite{www6}, and route planning~\cite{c17}. 
The problem of subgraph match querying entails the extraction of
subgraphs from an underlying graph having similar structure and labels to a
given query~\cite{www24, gfinder}.

Traditional approaches for exact structural and label matching based on
isomorphism are computationally infeasible. 
Thus, approaches based on pruning~\cite{www46,vf2,sg},
indexing~\cite{www36,www38}, filtering~\cite{www18,pbsm}, and dynamic
programming~\cite{dp} have been proposed. However, they fail to
scale for modern web-scale graph applications, wherein
{\em approximate
subgraph matching} was explored~\cite{www24,www27} to extract similar
subgraphs, with 
exact matches or 
with slight variations in
structural elements 
and label mismatches. For example, in
bioinformatics, approximate subgraph matching enables the detection of
candidate regions in genome, that might have undergone abnormal
mutations, for studying the associated medical effects~\cite{www37,www47}.
Although approximate subgraph extraction 
have been well 
studied~\cite{www24,www27,www38,www41,arora14,naga}, efficiently finding 
matching subgraphs with improved runtime and accuracy remains an important 
problem.

\noindent {\bf Problem Statement.} Consider $\newsym{G} = \left( \newsym{V_G},
\newsym{E_G}, \mathcal{L_G} \right)$ to be an input graph, where $\newsym{V_G}$
and $\newsym{E_G}$ denote the vertex and edge sets respectively, while
$\newsym{L_G}: \mathcal{V_G} \rightarrow \Gamma$ maps the vertices in
$\newsym{G}$ to a finite label (or attribute) set $\Gamma$.  A similar query
graph is considered: $\newsym{Q} = \left( \newsym{V_Q}, \newsym{E_Q},
\newsym{L_Q} \right)$. Without loss of generality, we assume that the graph $G$
is deterministic, undirected, vertex labeled, and does not contain hyper-edges.
The problem of {\em approximate subgraph matching} aims to find the top-k
subgraphs of \newsym{G} that are \emph{best} matching (maximum similarity) to
\newsym{Q} in terms of vertex label and edge overlap. In our context, \name
finds the \emph{top-k statistically significant subgraphs} of \newsym{G} as the
best approximate matches of \newsym{Q}.

\begin{figure}[t]
  \includegraphics[width=0.85\columnwidth]{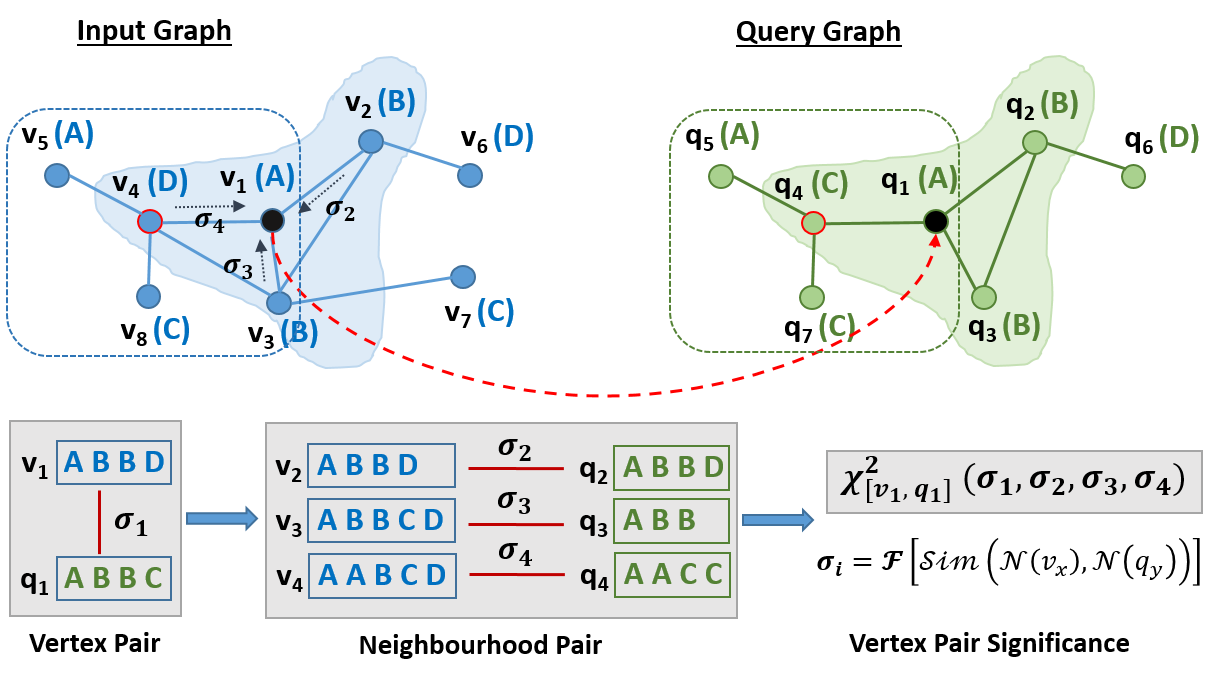}
  \caption{\small Two-hop neighborhood similarity based computation of
	$\chi^2$ statistical significance for vertex match in \name.}
  \label{fig:ex}
\end{figure}

\noindent {\bf State-of-the-art.} (Sub-)graph matching has been 
extensively studied, 
and the existing body of work can
be broadly categorized into two groups -- {\em exact} methods and {\em
approximate} heuristics. Since graph isomorphism 
is quasi-polynomial~\cite{www3} and subgraph isomorphism is
NP-complete~\cite{www10}, earlier works on exact graph matching 
such as Swift-Index~\cite{www36},
VF2~\cite{www11,vf2}, PathBlast~\cite{www23}, SAGA~\cite{www41},
IsoRank~\cite{www38} and GraphGrep~\cite{www18} to name a few, 
explored pruning and indexing techniques. 
To tackle
incomplete and noisy data, approximate matching techniques tolerate small
amounts of structural and label mismatches. These methods usually rely
on identifying candidate vertices, whose neighborhoods are then progressively
expanded in a greedy manner -- providing compute efficiency, although
with possibly sub-optimal results. Initial approaches like
TALE~\cite{www42}, C-Tree~\cite{www56}, GString~\cite{www22} and
SAPPER~\cite{www54} were based on indexing techniques and graph distance
measures to compute the degree of similarity. The use of pattern matching,
semantic-based search, and graph decomposition for finding matching subgraphs
were explored in gIndex~\cite{www50}, FG-Index~\cite{www8},
iGraph~\cite{www19}, Grafil~\cite{www51}, GPTree~\cite{www52},
cIndex~\cite{www7}, iGQ~\cite{www48}, and SIM-T~\cite{www27}. Surveys on the
rich literature of graph matching can be found in~\cite{www9,www17,sur1,sur2}.
Recent techniques like NeMa~\cite{www24} and GFinder~\cite{gfinder} adopt a
combination of efficient indexing and graph traversal based cost measure to
efficiently identify candidate matching regions, while VELSET~\cite{naga} and
NAGA~\cite{naga} use statistical measure to mine subgraphs that demonstrate 
significant deviations from the background distribution when matched to a query. 
Approximate graph matching in the context of probabilistic
graphs have also been studied~\cite{c45,chisel,c57,c63,chi80,c96,c33}.

\noindent {\bf Preliminaries.} {\em Statistical significance}
models the relationships between the observations and the factors that affect
the system. The p-value~\cite{pval} measures the probability of attributing an
observed event to chance or randomness. Since extreme events are rare, they
exhibit a smaller p-value or a higher statistical significance. Since,
computation of p-value is exponential, the {\em Pearson's chi-square statistic}
($\chi^2$) has been shown to provide an estimate~\cite{chi} of the
``goodness-of-fit'' of the set of observations. It is computed as the
normalized squared difference between the expected and observed occurrence
counts of the different outcomes.  Mathematically, $\chi^2 = \sum_{\forall
i}\left[\left(O_i - E_i\right)^2 / E_i \right]$, where $O_i$ and $E_i$ are the
observed and expected number of occurrences, respectively, for all outcomes
$i$. The {\em Chebyshev's inequality}~\cite{cheby} models the probability of
deviation for a random variable in terms of the number of standard deviations
from the distribution mean. Thus, for a random variable $X$ with finite mean
$\mu$ and non-zero variance $\delta^2$, we have $\Pr(| X - \mu | \geq t \cdot
\delta) \leq 1 / t^2$ for any $t>0$ ($t\in \mathbb{R}$). Intuitively, the
degree of label and structural overlap (i.e., similarity) between a query and
its matching subgraph would demonstrate significant deviations (due to high
similarity) from the expected characteristics (considering a random subgraph).
The Chebyshev's inequality can be used to characterize the \emph{difference} in
terms of the number of standard deviations away from the mean to compute the
statistical significance of candidate matching subgraphs. Such techniques have
been studied for sequence mining~\cite{substring,sachan,substr12}, substring
matching~\cite{mist}, subgraph similarity~\cite{naga,chisel}, and clique
finding~\cite{clique}.

\noindent {\bf Contributions.} In this paper, we propose the {\em
\underline{V}ertex N\underline{e}ighborhood Agg\underline{r}egation for
\underline{S}tatistically Signific\underline{a}nt Subgraphs via
\underline{Ch}ebyshev \underline{I}nequality} (\name) algorithm for efficient
top-k subgraph matching based on statistical significance. We
identify candidate neighborhood regions matching an input query by using
{\em two-hop} label and structural overlap based similarity. The deviation of the
observed similarity, from the underlying distribution
is then
characterized 
by 
Chebyshev's inequality and represented as symbols. Based 
on statistical significance, matching candidate regions are identified and explored 
in a greedy manner, to obtain the best matching subgraph to the query. Observe 
that \name adopts the methodology of~\cite{clique} for finding subgraph matches, 
while differing from~\cite{naga} in 
symbol computation and neighborhood
similarity. 
Initial empirical results on 
real and synthetic datasets 
showcase our proposed framework to outperform 
existing techniques in 
accuracy 
and 
robustness to noise.

\section{\name Algorithm}
\label{sec:algo}

This section describes the \name algorithm for extracting the top-k best
approximate matching subgraphs from a target graph $\newsym{G} = \left(
\newsym{V_G}, \newsym{E_G}, \newsym{L_G} \right)$, with respect to a query
graph $\newsym{Q} = \left( \newsym{V_Q}, \newsym{E_Q}, \newsym{L_Q} \right)$.
The working of \name comprises the following steps.  The first 5 steps are
\emph{offline} and are done only once for a target graph, while the last 4 are
\emph{online} and take place when a query arrives.

\noindent
{\bf 1. Index Creation.} Given a target graph \newsym{G}, \name initially
constructs two indexing lists summarizing the labels of the vertices and their
neighbors.  The first is an {\em inverted list}, \ilg, that maps vertex labels
to the corresponding list of vertices having the label. The second index, the
{\em label neighbor list}, \lnlg, stores the label information of the
neighbors for each vertex in \graph{G}.  A {\em label count vector} index,
$\lcvg(u)$, for each vertex $u$ in \graph{G} is also constructed.  It stores
the count of occurrence of each label (for $|\Gamma|$ labels) in the
neighborhood of $u$.  This enables efficient computation of similarity between
vertices as described next (step 4 onwards).

\noindent
{\bf 2. Similarity Measure.} For a vertex pair $(u,v)$ 
we use a modified {\em Tversky index}\footnote{$S(X,Y) = \frac{|X \cap Y|}{|X
\cap Y| + \alpha|X \setminus Y| + \beta|Y \setminus X|}$ for sets $X$ and $Y$
with parameters $\alpha, \beta \geq 0$.}~\cite{tindex} to define the {\em
vertex similarity score} ($\eta$):
{\small 
\begin{align}
	\label{eq:simi}
	\eta_{u,v} = |\mathcal{N}(u) \cap \mathcal{N}(v)| \big/ \left( |\mathcal{N}(u) \cap \mathcal{N}(v)| + |\mathcal{N}(v) \setminus \mathcal{N}(u)|^\gamma \right)
\end{align}
}
\noindent where $\mathcal{N}(u)$ is the set of labels in the neighborhood of $u$
including the label of $u$ itself.  Observe, by setting $\gamma=1$, we
obtain the original Tversky index with $\alpha = 0$ and $\beta = 1$.
Intuitively, the similarity of $u \in \graph{G}$ is maximized w.r.t. 
$v \in \graph{Q}$ when all neighbor labels of $v$ are present in the
neighbors of $u$ (i.e., $\mathcal{N}(v) \subseteq \mathcal{N}(u)$). Since the
presence of additional neighbors of $u \in \graph{G}$ should not affect the
similarity, we set $\alpha=0$ in the Tversky index.  In essence,
Eq.~\eqref{eq:simi} captures the {\em neighborhood recall} of $v \in \graph{Q}$
provided by $u \in \graph{G}$ (thus, $\beta=1$).  The exponential {\em
penalty factor}, $\gamma$, penalizes increasing mismatches in the
neighborhood label overlap between vertex pairs. It captures fine
differences in the neighborhoods by accentuating even smaller mismatches. 
Empirically, $\gamma=3$ gave the best results.

\noindent
{\bf 3. Initialization.} Using $\lcvg(u)$ structures and similarity measure,
\name computes the vertex similarity scores for every vertex pair in \graph{G}.
This captures the underlying distribution.  The expected similarity
distribution across random neighborhoods of \newsym{G} is captured via $3$
characteristics computed using $\eta_{u,w}: \forall \langle u \in \graph{G}, w
\in \graph{G} \rangle$. \\
(a) $\psi(\graph{G}) = {\sum_{u,w \in \mathcal{G}} \eta_{u,w}} /
|\graph{V_G}|:$ \emph{average} vertex similarity score for all vertex pairs in
\graph{G}, \\
(b) $\delta(\graph{G}) = \big(\sum_{u,w \in \mathcal{G}} (\eta_{u,w} -
\psi(\graph{G}))^2 / (|\mathcal{V_G}|-1)\big)^{1/2}:$ \emph{standard
deviation} of the vertex similarity scores of \graph{G}, and \\
(c) $\triangle(\graph{G}) = \max_{u,w \in \graph{G}}\{(|\eta_{u,w} -
\psi(\graph{G})| / \delta(\graph{G})\}:$ \emph{maximum} deviation of vertex
similarity score from the average among all the vertex pairs in terms of
standard deviations in \graph{G}.

\noindent
{\bf 4. Symbol Categorization.} The degree of matching between a target graph
vertex and a query vertex is captured in \name by the amount of deviation of
the vertex pair similarity score (in terms of the number of standard
deviations) from the underlying expected distribution (computed above).  The
standard deviations are discretized using the \emph{step size} parameter,
$\kappa$.  It also determines the total number of possible symbols,
$\tau = \lceil (\triangle(\graph{G}) - 1) / \kappa \rceil$.  The set of
category symbols, therefore, is $\Sigma = \{\sigma_1, \sigma_2, \cdots,
\sigma_\tau\}$.  Smaller values of $\kappa$ is preferred for differentiating
between finer-grained structural mismatches.

For a pair of vertices $u,w$, its similarity is characterized using the symbol
$\sigma_i$, $1 \leq i \leq \tau$.  The first symbol, $\sigma_1$, spans the
range of standard deviations up to $1 + \kappa$, i.e., $\sigma_1: 0 \leq
|\eta_{u,w} - \psi(\graph{G})| / \delta(\graph{G}) < 1 + \kappa$. Subsequent 
symbols cover step size standard deviations each, $\sigma_i: 1 +
(i-1) \cdot \kappa \leq |\eta_{u,w} - \psi(\graph{G})| / \delta(\graph{G}) < 1
+ i \cdot \kappa$ for $2 \leq i \leq \tau$.

\noindent
{\bf 5. Expected Probabilities of Symbols.} The expected probability of
occurrence associated with the category symbols is next computed using the {\em
Chebyshev's inequality}. Observe, the deviation of vertex pair similarity from 
the mean can be in negative or positive direction. 

Since we are interested in vertices that have higher
similarity than the mean (to capture higher matching), we only discretize the 
similarity (into symbols) when it is greater than the mean.  For all similarities 
that are lesser than the mean, we fold them into symbol $\sigma_1$. Thus, assuming 
symmetric one-sided Chebyshev's inequality, the occurrence probability of symbol $\sigma_i$ 
is $\Pr(\sigma_i) = \frac{1}{2} \left[ \frac{1}{\left( 1+ (i-1)
\cdot \kappa\right)^2} - \frac{1}{\left( 1+ i \cdot \kappa\right)^2} \right]$
for $2 \leq i \leq \tau$, and $\Pr(\sigma_1) = 1 - \sum_{j=2}^\tau
\Pr(\sigma_j)$.

We also empirically evaluated the variant where the deviation in both the positive and 
negative side of the mean are considered (i.e., without folding). However, it produced 
no changes in our results. Since a very low similarity (large negative deviation) can 
potentially have large chi-square values and, thus, produce false matching results, 
\name uses the one-sided version.

Note that all the above steps are {\em offline} 
operations and performed only once for a target graph \graph{G}.  

\comment{
In our setting, we use a value of $k=0.001$. \ab{we do
experiment with it}  Other tail bounds or domain-dependent probability
distributions might be used depending on the application, making \name robust
to diverse domains with different distributions \ab{probably this line can be
omitted, although I see the reason why it is mentioned -- for flexibility}.
}

\noindent
{\bf 6. Candidate Pair Mapping.} Upon arrival of a query graph \newsym{Q}, the
\emph{online} processing starts with the construction of indexes \ilq, \lnlq, and 
\lcvq, analogously to \graph{G}.  For each \emph{label} in \graph{Q}, \name creates 
{\em candidate pairs} between the vertices of \graph{G} and \graph{Q} having the 
same label. These candidate pairs form the initial seed vertex for extracting 
matching subgraphs (to the query) via greedy neighborhood search. Formally, the 
candidate pairs generated are $\vpm = \left\{\langle v \in \graph{G}, q \in \graph{Q}
\rangle ~|~ \mathcal{L_G}(v) = \mathcal{L_Q}(q)\right\}$.

\noindent
{\bf 7. Vertex Symbol Sequence.} For a candidate pair $\langle v
\in \graph{G}, q \in \graph{Q} \rangle$, \name computes the vertex pair
similarity score, $\eta(v,q)$, and characterizes the similarity score by
assigning a category symbol based on the deviation from the expected similarity
distribution (as discussed previously). The category symbol $\sigma_{\langle v,q \rangle}$  
captures the one-hop neighborhood similarity for vertices $v$ and $q$ (see Eq.~\eqref{eq:simi}).

We next compute ``second-order'' candidate pairs between the vertex sets adjacent to $v$ 
and $q$. A \emph{greedy} best mapping based on the vertex pair similarity score is used to 
compute the second-order candidate pairs. Similar to $\langle v, q \rangle$, each second-order 
candidate pair is assigned a category symbol based on the deviation of its similarity score
from the expected. The initial category symbol $\sigma_{\langle v,q \rangle}$ is aggregated 
with the second-order category symbols to form the {\em vertex symbol sequence}, 
$O_{\langle v,q \rangle}$, for the candidate pair $\langle v, q \rangle$. 

As an example, consider Fig.~\ref{fig:ex} depicting an initial candidate pair
between vertices $v_1$ and $q_1$ (both having label $A$) with category symbol
$\sigma_1$ assigned to it (using Eq.~\eqref{eq:simi}). The adjacent vertices of
$v_1$ ($\{v_2, v_3, v_4\}$) and the neighbors of $q_1$ ($\{q_2, q_3, q_4\}$)
are then greedily best-matched based on vertex pair similarity to obtain the
``second-order'' candidate pairs. For instance, $v_2$ and $q_2$ provides the
best match with the same label and the same neighborhood labels and, thus,
forms the next candidate pair (with, say, category symbol $\sigma_2$).
Subsequently, $v_3$ and $q_3$ are matched having the same label and partial
neighborhood overlap (consider to be assigned symbol $\sigma_3$). Finally, the
candidate pair $\langle v_4, q_4 \rangle$ is obtained with category symbol
$\sigma_4$. The corresponding vertex symbol sequence, $O_{\langle v_1,q_1
\rangle} = \{\sigma_1, \sigma_2, \sigma_3, \sigma_4\}$, associated to $\langle
v_1, q_1 \rangle$, captures the two-hop similarity between the candidate pair
vertices $v_1$ and $q_1$ (Fig.~\ref{fig:ex}).

\noindent
{\bf 8. Statistical Significance.} The computed symbol sequence $O_{\langle v,q
\rangle}$ signifies the degree of matching between the two-hop neighborhoods of
$v$ and $q$. Assuming $d$ to be the \emph{degree} of $q$, since mapping is
performed for the neighbors of $q$, the length of $O_{\langle v,q \rangle}$ is
$d$.  Thus, the {\em expected occurrence counts} of category symbol $\sigma_i$
is $E[\sigma_i] = d \cdot \Pr(\sigma_i)$.  The {\em observed occurrence counts}
of the category symbols are directly obtained from $O_{\langle v, q \rangle}$.
Using the observed and expected counts, \name computes the \emph{chi-square
statistics}, $\chi^2_{\langle v,q \rangle}$, for all the candidate pairs
obtained in \vpm (see step 6). \\
{\bf 9. Approximate Matching.} The candidate pairs along with their computed
chi-square values, $\langle v, q, \chi^2_{\langle v,q \rangle} \rangle$, are
inserted into a {\em primary max-heap} structure.  The candidate pair with the
largest $\chi^2$ value is extracted (assume $\langle v, q \rangle$) for
initializing the top-1 matching subgraph, $\match{1}$, and is considered as the
starting seed vertex for greedy expansion to find matching subgraph region for
the query \graph{Q}.

Next, candidate pairs between the adjacent vertices of the extracted seed pair
($v$ and $q$) are constructed (as in step 6) and pushed into a {\em secondary
max-heap} structure.  As before, the vertex symbol sequence of the candidate
pairs in the secondary heap are constructed, their statistical significances
computed, and the pair 
with the highest $\chi^2$ value is extracted and added to $\match{1}$. This
process is iterated till the secondary heap is empty, or the size of
$\match{1}$ equals the number of vertices in \graph{Q}. The subgraph obtained
in $\match{1}$ is reported as the {\em top-1 best approximate matching
subgraph} for \graph{Q}.

Vertices extracted from the primary and secondary heaps are marked as ``done''
to prevent duplicate entries in the heap structures, and ensuring that the same
region is not repeatedly explored. To retrieve more {\em top-k} approximate
matches for a query, the secondary heap is reset and the process is re-run,
starting from picking the currently best candidate pair (with highest
statistical significance) from the primary heap.  This is repeated until $k$
matches are obtained.

\begin{table*}[h!]
\centering
\small
\resizebox{0.96\textwidth}{!}{
	\subfloat[]{
		\begin{tabular}{cccc}
			\toprule
			{\bf Dataset} & {\bf \# Vertices} & {\bf \# Edges} & {\bf \# Unique Labels} \\
			\midrule
			{\tt Human} & 4,674 & 86,282 & 44 \\
			{\tt HPRD} & 9,460 & 37,081 & 307 \\
			{\tt Protein} & 43,471 & 81,044 & 3 \\
			{\tt Flickr} & 80,513 & 5.9M & 195 \\
			{\tt IMDb} & 428,440 & 1.7M & 22 \\
			\bottomrule
		\end{tabular}	
	}
	\subfloat[]{
		\begin{tabular}{c ccccc c ccccc}
			\toprule
			{\bf Dataset / } & \multicolumn{5}{c}{\bf Accuracy} && \multicolumn{5}{c}{\bf Running Time (sec)} \\
			\cline{2-6} \cline{8-12}
			{\bf Algorithm} & {\tt Human} & {\tt HPRD} & {\tt Protein} & {\tt Flickr} & {\tt IMDb} && {\tt Human} & {\tt HPRD} & {\tt Protein} & {\tt Flickr} & {\tt IMDb} \\
			\midrule
			{\em VELSET}~\cite{naga} & 0.42 & 0.65 & 0.37 & 0.75 & 0.53	&& 0.01 & 0.01 & 0.16 & 0.13 & 1.31	\\
			{\em G-Finder}~\cite{gfinder} & 0.45 & 0.12 & 0.47 & \multicolumn{2}{c}{\ \ out of memory} && 0.55 & 0.01 & 0.12 & \multicolumn{2}{c} {\ \ \ out of memory} \\
			{\em \name} & \bf 0.90 & \bf 0.81 & \bf 0.67 & \bf 0.84 & \bf 0.87 && 0.12 & 0.06 & 0.77 & 1.98 & 6.90 \\
			\bottomrule
		\end{tabular}
	}
}
\caption{\small (a) Summary of the datasets characteristics. (b) Overall accuracy and runtime performance of the algorithms on the different datasets.}
\label{tab:data}
\end{table*}

\begin{figure*}[!htb]
    \centering
    \begin{minipage}{0.67\textwidth}
        \centering
		\begin{tabular}{cc}
		\includegraphics[width=0.45\columnwidth]{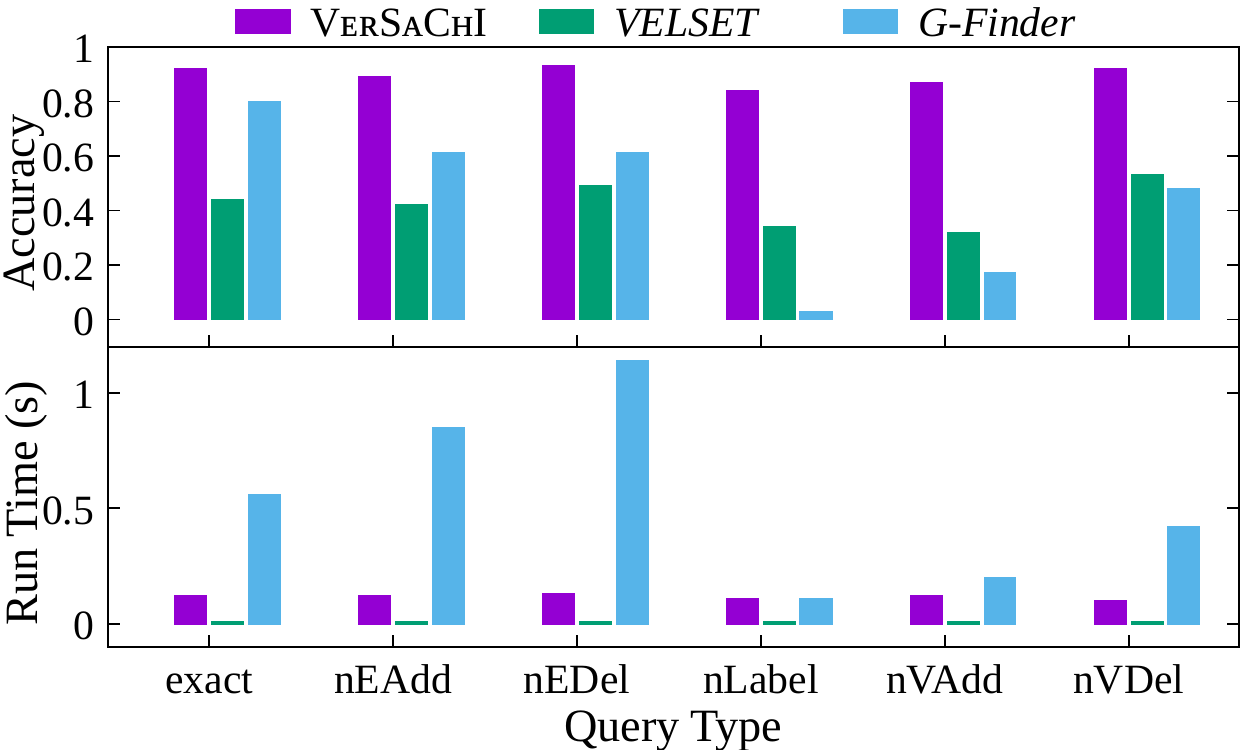} & 
		\includegraphics[width=0.45\columnwidth]{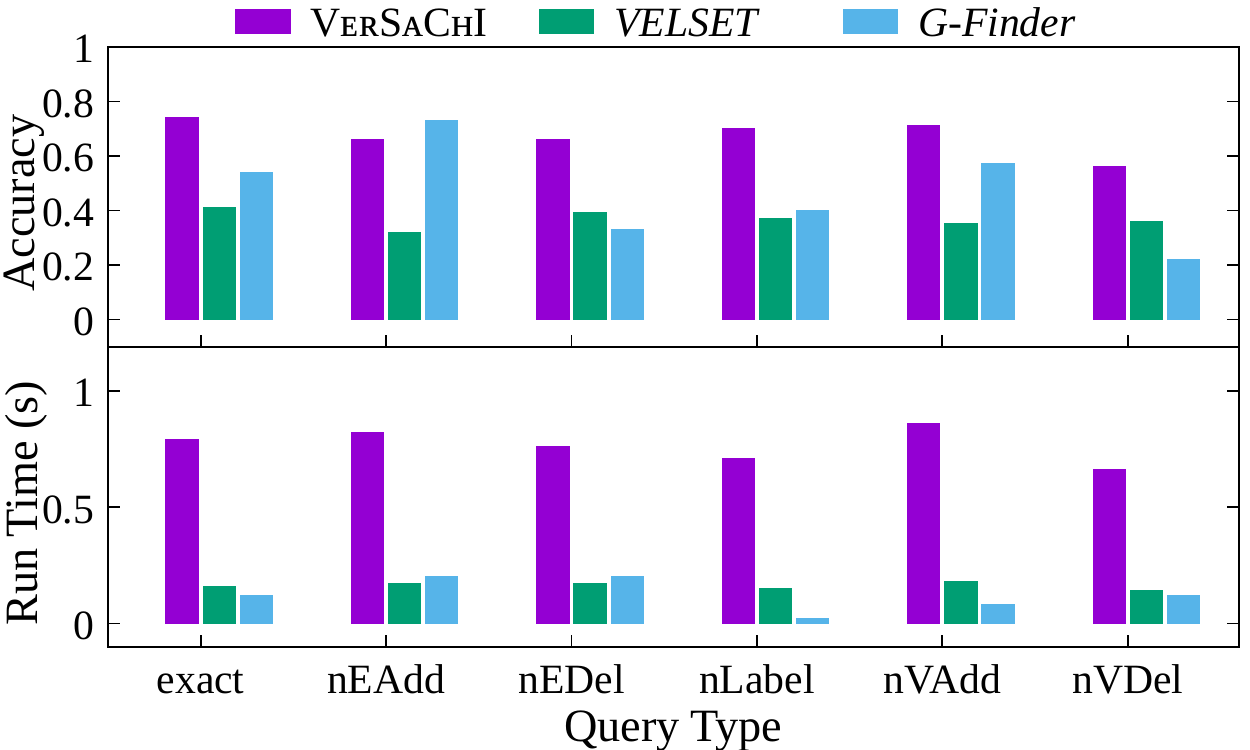} \\
			{\bf (a)} {\tt Human} & {\bf (b)} {\tt IMDb}
		\end{tabular}
        \caption{Performance for different query types on {\tt Human} and {\tt IMDb} datasets.}
        \label{fig:qtype}
    \end{minipage}%
    \begin{minipage}{0.3\textwidth}
        \centering
		\hspace*{-1mm}\includegraphics[width=1.05\columnwidth]{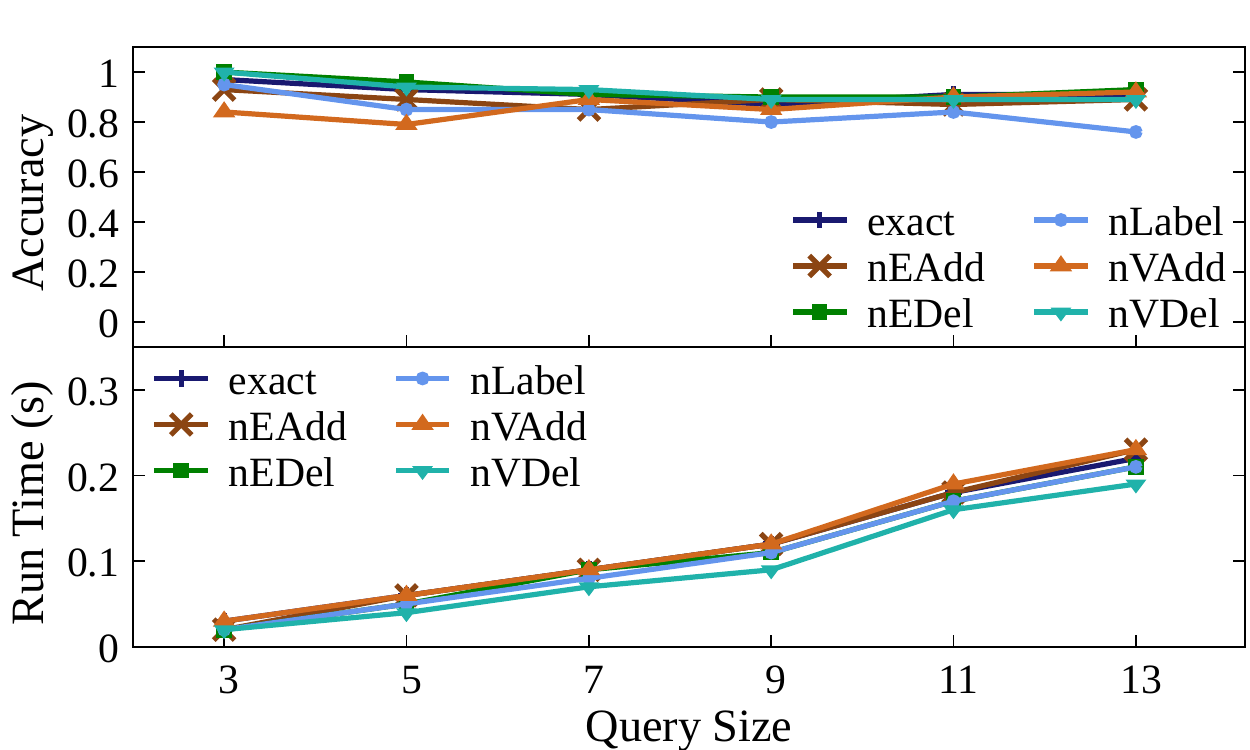}
		\caption{Performance with query size on {\tt Human} dataset.}
        \label{fig:qsize}
    \end{minipage}
\end{figure*}

\begin{figure*}[h]
\centering
\resizebox{\textwidth}{!}{
	\begin{tabular}{ccc}
	\includegraphics[width=0.33\textwidth]{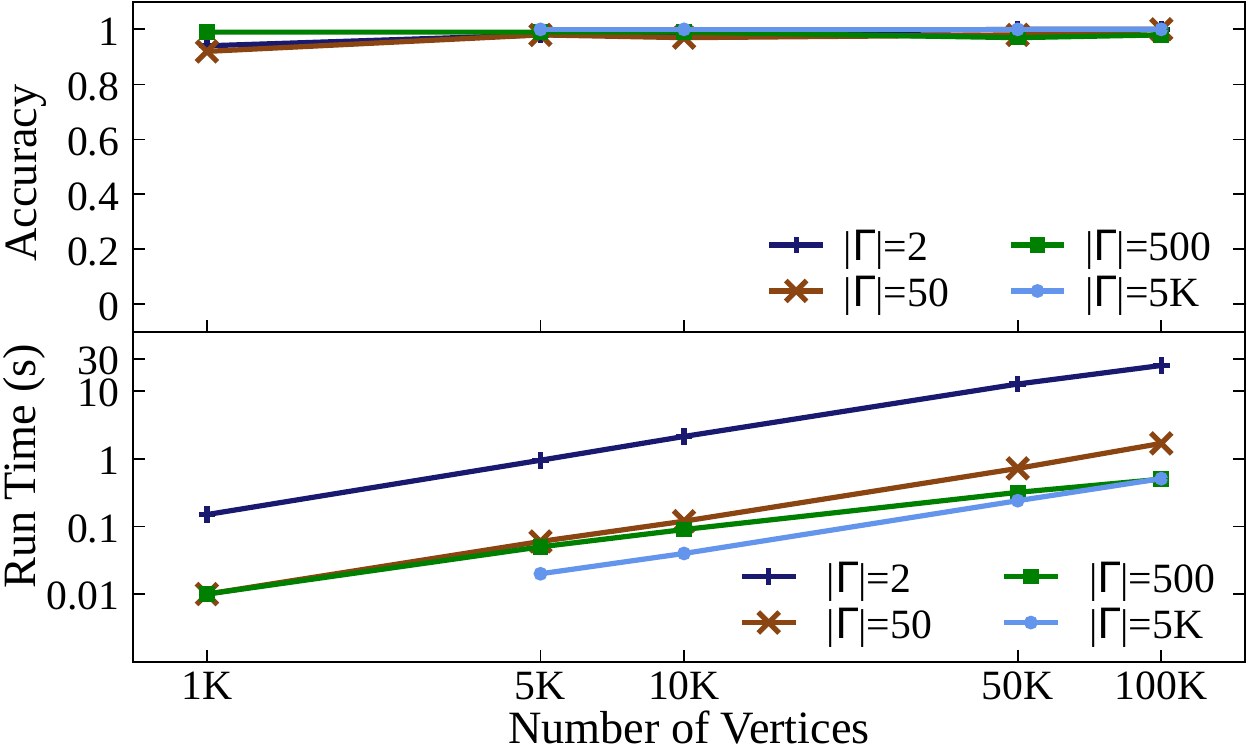} & 
	\includegraphics[width=0.33\textwidth]{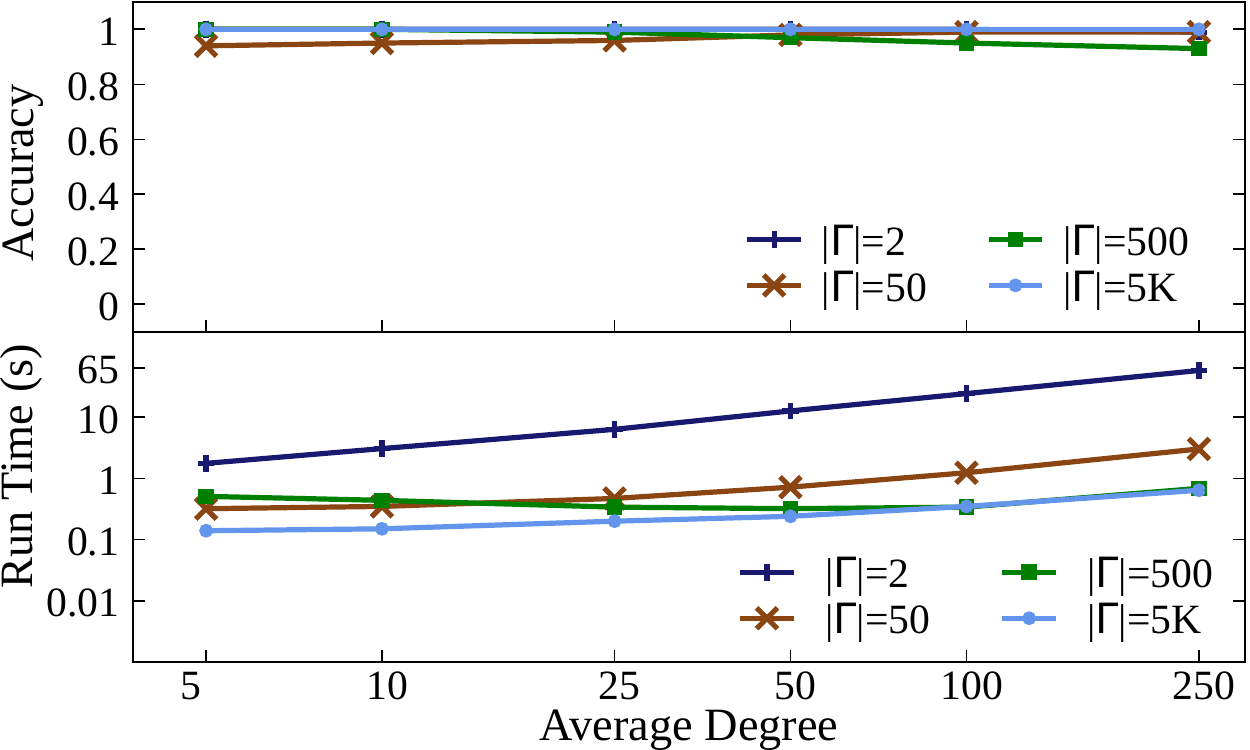} & 
	\includegraphics[width=0.33\textwidth]{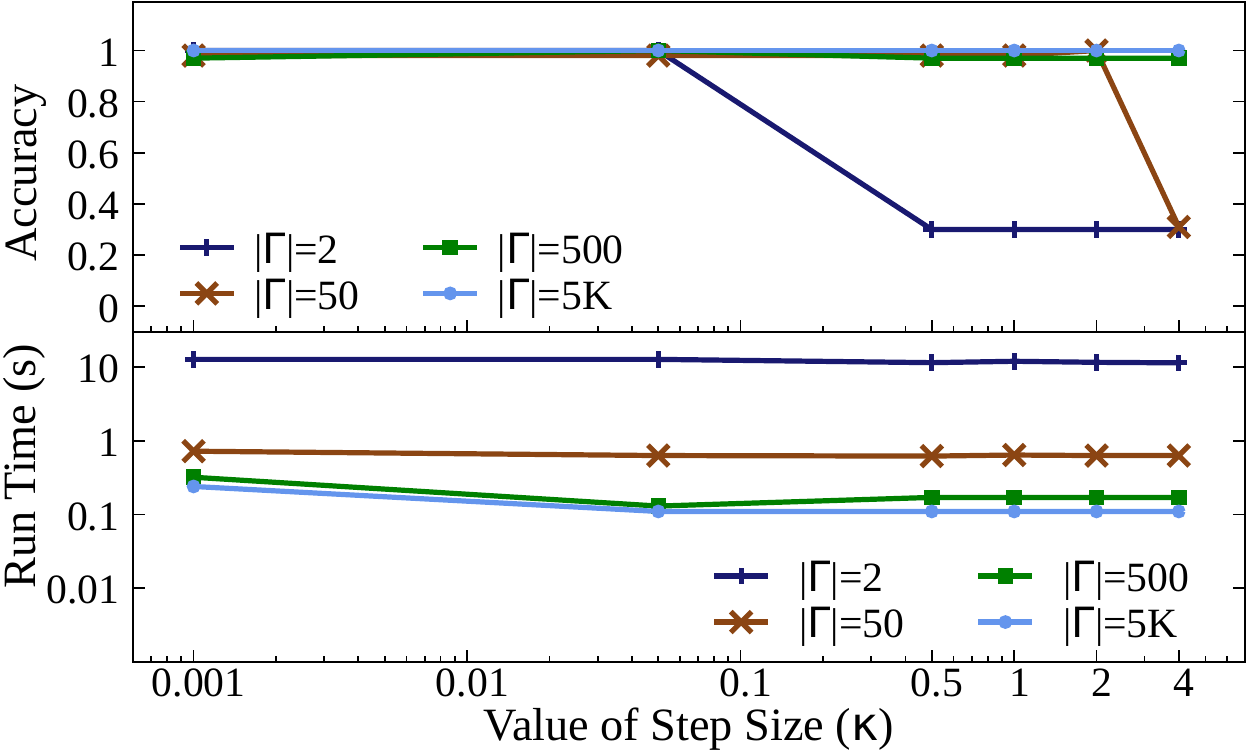} \\
	{\bf (a) Number of Vertices (Avg. Degree = 50, $\kappa$ = 0.001)} &
	{\bf (b) Average Degree (|V| = 50K, $\kappa$ = 0.001)} &
	{\bf (c) Step Size ($\kappa$) (|V| = 50K, Avg. Degree = 50)}\\
	\end{tabular} 
}
	\caption{\small Performance of \name on Barab\'asi-Albert graphs with varying (a) number of vertices ($n$), (b) average degree, and (c) step size ($\kappa$).}
	\label{fig:scalability}
\end{figure*}

\noindent {\bf Complexity Analysis} 

Assume graph \graph{G} to contain $n$ vertices, $m$ edges and $|\Gamma|$ unique 
labels. Index construction (offline phase) requires $O(n)$ space for \ilg, $O(m)$ 
for \lnlg, $O(n \cdot |\Gamma|)$ for \lcvg, and $O(\tau)$ for symbol probabilities. 
The overall space complexity of \name, therefore, is $O(n + m + \tau + n \cdot |\Gamma|)$. 
The time taken for index construction are $O(n)$ for \ilg, and $O(m)$ for both
\lnlg and \lcvg. Computing the target graph underlying distribution requires
traversal of \lcvg for each vertex pair in \graph{G}. Thus, total
\emph{offline} time is $O(n + m + n^2 \cdot |\Gamma|) \approx O(n^2)$.

Once a query arrives, for each query vertex, candidate pairs (with same
label) are constructed using the inverted indices. Assuming uniform
label distribution in \graph{G}, the number of candidate pairs is
$O(n_\mathcal{Q} \cdot n / |\Gamma|$). For each candidate pair, vertex
symbol sequence generation (both initial and ``second-order'') takes $O(\rho
\cdot |\Gamma|$) time where $\rho$ is the maximum degree in \graph{G}. Since
$\chi^2$ computation takes $O(\tau)$ time, the overall runtime of
\name is $O(n_\graph{Q} \cdot n / |\Gamma|)$.

\section{Experiments}
\label{sec:exper}

In this section, we discuss the empirical setup and evaluation of the \name
algorithm, and its comparison to existing approaches.

\noindent
{\bf Datasets.} We evaluate the performance of the algorithms on real datasets
from $3$ different domains: (i)~{\it Biological Networks:} protein-protein
interaction graphs of {\small \tt Human}, {\small \tt HPRD}~\cite{data} and
{\small \tt Protein}~\cite{repo}; (ii)~{\it Social Interaction:} social
interaction network between users of the image and video hosting site Flickr,
with the label of each user (vertex) denoting the group that she belongs
to~\cite{repo}; and (iii)~{\it Knowledge Graph:} {\small \tt IMDb}~\cite{repo}
containing named-entities like movies, actors, etc., along with their
relationships. The characteristics of the datasets are shown in
Table~\ref{tab:data}(a). Synthetically generated {\em Barab\'asi-Albert} graphs
are also used to study the scalability of \name.

\noindent
{\bf Query.} Query graphs (connected) are constructed (from the dataset) by
initially selecting a random vertex, and exploring its neighborhood till
$n_\graph{Q}$ vertices are visited. These are referred to as {\em exact
queries}. To study the performance of the algorithms in presence of noise,
exact query graphs were perturbed by introducing structural and label noise
randomly by (i)~modifying vertex labels ({\em nLabel}), or (ii)~inserting
or deleting vertices ({\em nVAdd} and {\em nVDel} resp.), or
(iii)~adding or deleting edges ({\em nEAdd} and {\em nEDel}
resp.). The number of perturbations are limited to $2$.
Further, for each scenario, we generate queries with sizes varying from $3$ to
$13$ (at intervals of $2$), with $20$ query graphs extracted for each size.
Thus, for each dataset, we consider ($6 \times 6 \times 20$) = $720$ queries,
and report average results.

\noindent
{\bf Evaluation.} The efficiency of the algorithms are measured in terms of
{edge retrieval accuracy} (using the labels of end vertices), that is, the
fraction of edges of the query graph \newsym{Q} that are present in the
matching subgraph retrieved. Additionally, we report the average runtime
required (per query) by the different approaches to extract the approximate
matching subgraphs.  Since the introduced perturbations do not exist in the
original graph, the exact query (for obtaining the noisy query) is considered
as the ground truth. For {\em Barab\'asi-Albert} graphs we use exact queries only.

\noindent
{\bf Baselines.} We compare the performance of \name algorithm
against the following: (i) {\em VELSET}~\cite{naga}, a statistical
significance based approach for exploring candidate regions with partial label
match;
and (ii) {\em G-Finder}~\cite{gfinder}, a
graph traversal based indexing for dynamic filtering and refinement of
candidate match neighborhoods. \\
%
{\bf Index.} The maximum index size taken by \name in our experiments is $1.4$\,GB for the
{\tt Flickr} graph, while the highest offline computation time is 32783.44 seconds, for 
{\tt IMDb} dataset.

\noindent
{\bf Setup.} All experiments were implemented in C++ and were conducted on an
Intel(R) Xeon(R) 2.60GHz CPU E5-2697v3 with 500GB of RAM. G-Finder was obtained
from {\small \url{github.com/lihuiliullh/GFinder}} and evaluated on a Visual
Studio 2015 C++ platform.

\noindent {\tt {\bf Empirical Results}}

From Table~\ref{tab:data}(b), we observe that \name has a significantly better
accuracy than the competing algorithms for finding the best matching subgraphs
with more than {\bf $20\%$} accuracy improvements (averaged across varying
query types and sizes). The run-time of \name is slightly more than the other
approaches due to the two-hop neighborhood similarity computation.  In absolute
terms, though, it is quite practical.  Overall, with a slight increase in
compute time, \name offers a substantial accuracy gain.  ({\em G-Finder}
crashes due to out-of-memory issues for {\tt Flickr} and {\tt IMDb} datasets.)

Fig.~\ref{fig:qtype} depicts the performance for different query types. (Results 
on the other datasets are similar and are, thus, omitted due to space constraints). 
\name achieves better accuracy
across all the different query types, with slight increase in runtime.
Fig.~\ref{fig:qsize} shows that with increase in query size, the runtime increases 
linearly (across query types), while the accuracy remains
largely unaffected.

Fig.~\ref{fig:scalability} studies the scalability of \name using synthetic
{\em Barab\'asi-Albert} graphs. The runtime is seen to increase linearly, with
increase in number of vertices and average degree, conforming to the analysis
of Sec.~\ref{sec:algo}. The accuracy of \name is unaffected in
these scenarios.  With increase in the {\em step size} $\kappa$, accuracy
decreases, as the number of symbols decreases, limiting the
power of \name to differentiate between finer differences in neighborhood
mismatches between the graphs, while the runtime remains mostly
constant.

\section{Conclusions}
\label{sec:conc}

This paper proposed a scalable and highly accurate algorithm, \name, for approximate labeled graph querying.
It shows significantly better accuracy than the competing methods across datasets and noise.
Our framework is generic enough
to accommodate other similarity measures and application-dependent tail distributions.

\balance
\bibliographystyle{ACM-Reference-Format}
\bibliography{biblio}

\end{document}